\documentclass[a4paper,10pt,twocolumn]{article}
\usepackage[utf8]{inputenc}
\usepackage{multicol,graphicx}
\usepackage{mathptmx}
\usepackage{bm} 
\usepackage[T1]{fontenc}
\usepackage{textcomp}
\usepackage{url}
\usepackage[hidelinks]{hyperref}
\usepackage[T1]{fontenc} 
\usepackage[english]{babel}
\usepackage{booktabs}
\usepackage{multirow}

\usepackage[backend=bibtex,style=ieee,doi=false,isbn=false,url=true,maxnames=6,citestyle=numeric-comp,giveninits=true]{biblatex}

\fontfamily{ptm}\selectfont

\bibliography{references.bib}

\usepackage[font=small]{caption}
\usepackage{amsmath, amssymb}

\newcommand{\mysection}[1]{\vspace{0.4cm} \uppercase{#1} \vspace{0.4cm}}
\newcommand{\mysubsection}[1]{\hspace{10pt}\textit{#1:}}

\begin{document}
	
\setlength{\textfloatsep}{10pt plus 1.0pt minus 2.0pt}	
\setlength{\columnsep}{1cm}


\twocolumn[%
\begin{@twocolumnfalse}
\begin{center}
	{\fontsize{14}{18}\selectfont
        \textbf{\uppercase{Towards auditory attention decoding with noise-tagging: a pilot study}}\\}
    \begin{large}
        \vspace{0.6cm}
        H. A. Scheppink\textsuperscript{1}, S. Ahmadi\textsuperscript{1}, P. Desain\textsuperscript{1}, M. Tangermann\textsuperscript{1}, J. Thielen\textsuperscript{1}\\
        \vspace{0.6cm}
        \textsuperscript{1}Donders Institute, Radboud University, Nijmegen, the Netherlands\\
        \vspace{0.5cm}
        E-mail: \href{mailto:jordy.thielen@donders.ru.nl}{jordy.thielen@donders.ru.nl}
        \vspace{0.4cm}
    \end{large}
\end{center}	
\end{@twocolumnfalse}%
]%


ABSTRACT: 
Auditory attention decoding (AAD) aims to extract from brain activity the attended speaker amidst candidate speakers, offering promising applications for neuro-steered hearing devices and brain-computer interfacing. This pilot study makes a first step towards AAD using the noise-tagging stimulus protocol, which evokes reliable code-modulated evoked potentials, but is minimally explored in the auditory modality. Participants were sequentially presented with two Dutch speech stimuli that were amplitude-modulated with a unique binary pseudo-random noise-code, effectively tagging these with additional decodable information. We compared the decoding of unmodulated audio against audio modulated with various modulation depths, and a conventional AAD method against a standard method to decode noise-codes. Our pilot study revealed higher performances for the conventional method with 70 to 100 percent modulation depths compared to unmodulated audio. The noise-code decoder did not further improve these results. These fundamental insights highlight the potential of integrating noise-codes in speech to enhance auditory speaker detection when multiple speakers are presented simultaneously.


\mysection{introduction}

People suffering from hearing loss often have great difficulty in scenarios in which multiple individuals are speaking simultaneously, known as the `cocktail party scenario', something which normal hearing persons have no difficulties with~\cite{cherry1953}. In these scenarios, hearing aids are not able to provide a good solution, as even though they are capable to suppress background noise, they are less capable of suppressing the unattended speakers. Some hearing aids attempt to mitigate this problem by using a heuristic, for example by enhancing the loudest or closest speaker, or the one who stands right in front of the listener. Unfortunately, these heuristics often lead to selecting the wrong speaker in real-life scenarios.

As an alternative, auditory attention decoding (AAD) aims to decode the attended speaker from neural activity, as the synchronization between the listener's brain signals and the attended speech envelope is stronger than with the ignored speech envelope~\cite{ding2012}. This finding laid the groundwork for more research on hearing aids that allow for cognitive control, so called neuro-steered hearing aids~\cite{geirnaert2021}. These hearing aids aim to identify the attended speaker from neural activity, and correspondingly enhance this speaker's audio signal whilst simultaneously suppressing the other speakers and background noise. 

The main idea behind such AAD approaches is to match the speech signals to the neural activity which synchronizes with the attended speech signal. For practical reasons typically electroencephalography (EEG) is used. Most AAD algorithms follow a \textit{stimulus reconstruction} approach, also referred to as \textit{backward modeling} or \textit{decoding}. In this approach, the attended speech envelope is reconstructed from the EEG using a neural decoder, before the speaker whose envelope has the highest similarity with the reconstructed envelope is assumed to be the attended speaker~\cite{osullivan2015}. Another approach is \textit{forward modeling} or \textit{encoding}, in which the objective is to predict the neural response from the speech envelopes via an encoder, and to compare these against the EEG~\cite{lalor2010, ding2012b}. A third approach, sometimes referred to as the \textit{hybrid} approach, combines decoding and encoding, by transforming both the speech envelopes and the EEG to minimize the irrelevant variance~\cite{cheveigne2018, dmochowski2018}. Using such a hybrid approach, Geirnaert and colleagues~\cite{geirnaert2021} achieved a remarkable performance using canonical correlation analysis (CCA) to decode the attended speaker. Presenting audio from two simultaneous speakers, a mean accuracy of about 85\,\% was reached using 30\,s decision windows. However, when decreasing the decision window length to about 10\,s, the accuracy quickly dropped to below 80\,\%. This poses a significant limitation for real-world scenarios where fast speaker detection is crucial.

Framing the speaker decoding problem as detecting which of several stimuli a person is attending to, another paradigm from the brain-computer interfacing (BCI) field recently reached remarkable performances. Specifically, the locus of visual attention can be decoded from EEG data using the code-modulated visual evoked potential (c-VEP). A c-VEP is the EEG response to pseudo-random visual stimulation sequences where stimuli are watermarked using noise-codes, a protocol called noise-tagging~\cite{martinez2021}. These noise-codes are selected or even optimized to be dissimilar, such that attending to one noise-code evokes substantially different brain activity than when attending to another code, facilitating the decoding of the attended stimulus. Such c-VEP BCIs have been reaching state-of-the-art performances up to 100\,\% classification accuracy using 1--4\,s decision windows~\cite{thielen2021} or recently even within 300\,ms~\cite{shi2024} and a high number of stimuli, 29 and 40, respectively.

This study aims to create fundamental insights in the application of the noise-tagging protocol for auditory attention decoding. This can be realized through ``watermarking'' the speech signal with the pseudo-random noise-codes. To accomplish this, we propose modulating the amplitude of each speech audio with the amplitude of a unique noise-code, effectively embedding the noise information within the speech signal. In turn, not only can we decode the attended speaker based on the speech envelope, but we can also leverage the hidden noise-tags for enhanced speaker identification. 

In this pilot study, we make the first step towards auditory attention decoding using noise-tagging. Firstly, we aim to assess the feasibility of decoding the code-modulated auditory evoked potential (c-AEP), the response to auditory noise-tagging. Therefor, we use sequential presentation, i.e., only one stimulus is presented at a time. Secondly, in this work we compare various modulation depths against no modulation. Thirdly, we compare how decoding approaches based on the speech envelope and noise-tag compare in terms of classification accuracy.

Successful implementation could make the step towards improving the decoding accuracy and speed in identifying the attended speaker. Furthermore, this exploration may pioneer a novel research avenue for the application of code-modulated responses in the auditory domain, a domain that has seen limited application compared to the visual modality, as so far only one study attempted this~\cite{farquhar2008}.

\mysection{materials and methods}

\mysubsection{Participants}
Five participants (aged 19--31 years, average 23 years, 3 females and 2 males) participated in the pilot experiment. Two of these participants were authors of this study. All participants gave written informed consent prior to the experiment. The experimental procedure and methods were approved by and performed in accordance with the guidelines of the local ethical committee of the Faculty of Social Sciences of Radboud University.

\mysubsection{Materials}
The EEG data were recorded at a sample rate of 500\,Hz with 64 active electrodes placed according to the 10-10 system and amplified by a BrainAmp (Brain Products GmbH) amplifier. The EEG data were preprocessed with a non-causal FIR notch filter at 50\,Hz and a bandpass filterbetween 1 and 20\,Hz before resampling to 120\,Hz.For filtering, the default settings were used from the MNE toolbox, version 1.6.1~\cite{Gramfort2013a}.

The auditory stimuli were two Dutch short stories~\cite{radioboeken}, narrated by two different male speakers and recorded at 44100\,Hz. They lasted approximately 6.5\,min each and were a subset of the stimulus materials used by Das and colleagues~\cite{das2016}. Periods of silence exceeding 500\,ms were truncated to 500\,ms. The stimuli were normalized for loudness and presented dichotically to participants via headphones, with one story consistently presented to the left and the other to the right ear.

We used two 126-bit binary pseudo-random noise-codes from a family of modulated Gold codes~\cite{gold1967, thielen2015} to amplitude-modulate the audio. The codes come in sets that are maximally uncorrelated with each other and each time-shifted versions of themselves. The codes were modulated to include only short (`010') and long (`0110') events. From the available modulated Gold codes, we carefully selected one that started with a 1, ended with a 0, and exhibited an almost uniform distribution of short and long events. The second code was a 61-bit phase-shifted version of the first. In this way, the noise-codes had identical properties, while minimizing auto-correlation at a maximum delay. The codes were presented at a bit rate of 40\,Hz, and we always used the first code to modulate audio for the left ear, while the second code was always used to modulate audio for the right ear.

\begin{figure*}[h]
    \centering
    \includegraphics[width=\linewidth]{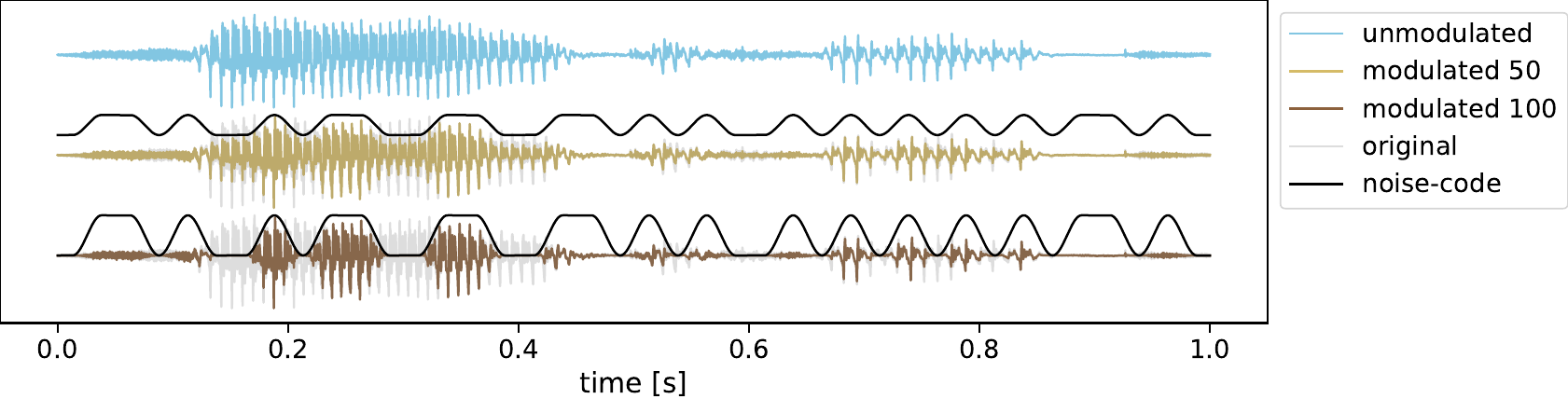}
    \caption{\label{fig:modulation}\textbf{Visualization of three different modulation depths using noise-tagging.} Depicted is the unmodulated audio, i.e., 0 percent (blue), and 50 (gold) and 100 (brown) percent modulated audio. Additionally, shown are the smoothened noise-tags used for modulation (black). Audio was amplitude-modulated by multiplying with the noise-code, retaining full audio amplitude when the code is 1, while only a percentage when it is zero. Therefore, the noise-code for 50 percent modulation ranges between 0.5--1, instead of 0--1 for 100 percent modulation. To ease comparison, we added the original audio (light gray) at the back of the modulated audio. }
\end{figure*}

The two stories were presented in their original form, or subjected to amplitude modulation using the binary noise-codes, as shown in Fig.~\ref{fig:modulation}. At full modulation depth, i.e. 100 percent, further denoted as condition 100, the audio was directly multiplied with the bit sequence, resulting in undisturbed audio when the noise-code is 1 and complete audio suppression when the code is 0. To avoid potential speech unintelligibility, smaller modulation depths were also tested. For instance, at modulation 90, the audio was dampened by 90\,\% when the code is 0 and otherwise retained. We tested five modulation conditions: 100, 90, 70, 50, and 0. In other words, condition 0 denotes the unmodulated audio. To address the abrupt bit transitions of the binary codes, we smoothened their edges with a raised cosine function, see Fig.~\ref{fig:modulation}.

\mysubsection{Experiment}
During the experiment, participants completed five runs, each corresponding to a distinct modulation condition. The order of conditions was randomized over participants. During a run, two trials were presented, one for each of the two stories, starting with the first story delivered to the left ear accompanied by silence on the right, followed by the second story presented to the right ear with silence on the left. The stories were presented sequentially, to assess the feasibility of auditory noise-tagging before testing the more complex parallel case, where stories would be presented simultaneously.

Throughout a run, a fixation cross was displayed on a mean-luminance gray background at the center of a monitor, positioned approximately 70\,cm in front of the participant. Each run started with a 5\,s rest, followed by a 1\,s cue that indicated the to-be attended side, succeeded by the audio presentation. Participants could take self-paced breaks between runs.

\mysubsection{Analysis}
The classification of the attended speech in this work was done using two approaches, the conventional envelope CCA (further denoted as eCCA) in which only the information of the envelope of the speech signals is used, and reconvolution CCA (further denoted as rCCA), in which both the envelope as well as the noise-codes is used. Both approaches can be applied for the modulated audio and corresponding data, however only the eCCA approach can be applied on the unmodulated audio condition as here no code information is available for rCCA. The eCCA is based on the implementation by Geirnaert and colleagues~\cite{geirnaert2021}. The use and implementation of rCCA follows the work by Thielen and colleagues~\cite{thielen2021}.

For demonstrating full selective attention, a parallel presentation of stimuli would be required. However, in this pilot study we aimed to make the first step towards auditory noise-tagging, hence the stimuli were presented sequentially. Specifically, when a stimulus was presented to the left ear there was no stimulus presented to the right ear, and vice versa. To evaluate the models, we did however simulate as if the other stimulus of the same condition was presented to the unattended ear, leading to a two-class problem. 

Both methods make use of envelope information, which were obtained as follows: the modulated and unmodulated speech signals were each first filtered by a gammatone filterbank resulting in 15 frequency subbands for each speech stimulus~\cite{biesmans2017}. Then, for each subband the absolute value was taken, followed by a power-law compression of 0.6. Next, each subband was lowpass filtered at 20\,Hz and resampled to 120\,Hz, which were a factor and multiple of the 40\,Hz bitrate of the noise-code, respectively. Finally, for each stimulus all subbands were summed, with equal weights, to obtain the stimulus-specific envelopes. 


For model evaluation within a single condition, we created four partitions of the data. Specifically, we divided the two 6.5-minute trials into four chronological segments and allocated one segment of each trial to one partition. These four partitions were used for 4-fold cross-validation. This resulted in training and test sets of 9.45\,min and 3.15\,min, respectively, that had a balanced label distribution. 

During the testing of a particular model for a condition, a sliding window was moved over the test data of one class with a stride of 2 samples. To test the sensitivity to data availability, we tested increasing decision window lengths of $\tau$ 1, 2, 5, 10, 20, 30 and 60\,s. The analysis yielded results per sliding window and folds, both of which were averaged to obtain the accuracy of one model in one condition at one decision window length for each participant.

In the following two sections, we explain the two decoding approaches, starting with the eCCA method followed by the rCCA.

\mysubsection{Envelope CCA}
The eCCA approach aims to find a direct correspondence between the EEG data and the speech envelope to detect the attended speaker. Let's assume EEG data $\mathbf{X} \in \mathbb{R}^{C \times T}$ of $C$-many channels (here \mbox{$C=64$}) and $T$-many samples (here $T$ is one segment size). Additionally, let's assume the speech envelope \mbox{$\mathbf{A}_i \in \mathbb{R}^{L \times T}$} for the $i$th speaker with $L$ time-lagged envelopes (here $L=60$ for 500\,ms at 120\,Hz) of $T$-many samples each. Then, eCCA optimizes a spatial filter $\mathbf{w} \in \mathbb{R}^C$ and a temporal filter $\mathbf{r} \in \mathbb{R}^L$ such that the projected data and envelope are maximally correlated. Let's assume we have a labeled training dataset $\{(\mathbf{X}_1, y_1), (\mathbf{X}_j, y_j), \dots, (\mathbf{X}_J, y_J)\}$ with $J$ segments. Then, CCA optimizes the following correlation $\rho$:
\begin{equation}\label{eq:ecca_fit}
    \underset{\mathbf{w}, \mathbf{r}}{\arg\max}~\rho (\mathbf{w}^{\top}\mathbf{S}, \mathbf{r}^{\top}\mathbf{Z}) 
\end{equation} 
where $\mathbf{S} = [\mathbf{X}_1, \mathbf{X}_j, \dots, \mathbf{X}_J]$ are the concatenated training EEG segments, and $\mathbf{Z} = [\mathbf{A}_{y_1}, \mathbf{A}_{y_j}, \dots, \mathbf{A}_{y_J}]$ are the accompanying concatenated speech envelopes.


To classify new data $\mathbf{X} \in \mathbb{R}^{C \times T}$ (here $T=\tau$ the decision window length), eCCA chooses the candidate speech envelope that maximizes the correlation $\rho$ between the spatially filtered EEG data and the projected speech envelopes:
\begin{equation}\label{eq:ecca_predict}
    \hat{y} = \underset{i}{\arg\max}~\rho(\mathbf{w}^\top\mathbf{X}, \mathbf{r}^\top\mathbf{A}_i)
\end{equation}

Instead of using the first component only, as in Eq.~\ref{eq:ecca_predict}, using multiple components can improve classification accuracy but requires an additional classification model, e.g., a linear discriminant analysis (LDA)~\cite{geirnaert2021}. Specifically, CCA can deliver $K=\min(C, L)$ orthogonal components, ordered on decreasing canonical correlation. The $k$-th component contributes a spatial filter $\mathbf{w}_k$ and temporal filter $\mathbf{r}_k$, and delivers a Pearson's correlation coefficient $\rho_{ki}$ following Eq.~\ref{eq:ecca_predict}. These correlation coefficients across $K$ components (here $K=3$) are collected in a vector ${\bm\rho}_i$ for speaker $i$, and a feature vector $\mathbf{f}$ is created by subtracting the speakers' canonical correlation vectors, $\mathbf{f} = {\bm\rho}_1 - {\bm\rho}_2$. The low-dimensional feature vector $\mathbf{f}$ can then be classified using a vanilla LDA, solving a binary classification problem of whether speaker 1 or speaker 2 was attended.

\mysubsection{Reconvolution CCA}
The rCCA approach consists of a template-matching classifier that predicts the attended speaker given the neural response evoked by the binary noise-code. The reconvolution model is based on the superposition hypothesis, stating that the response to a sequence of events is the linear summation of the responses evoked by the individual events~\cite{thielen2015}. 

For the reconvolution, the event time-series $\mathbf{E}_i \in \mathbb{R}^{E \times T}$ for $E$-many events and $T$-many samples (here $T$ is one segment size) denotes the onsets of the $E$ events for the $i$th noise-code. In this work, we used $E=2$ events being the short and long events in the noise-codes.

The event matrix is mapped to a structure matrix $\mathbf{M}_i \in \mathbb{R}^{M \times T}$ for $M$-many event time points and $T$-many samples. This matrix maps each event to an impulse response function. Specifically, this matrix is Toeplitz-like and describes the onset, duration, and importantly the overlap of each of the events. Assuming both events evoke a response of identical length $L$, then $M=E*L$ (here $L=60$ for 500\,ms at 120\,Hz).

In this work, we extend the standard rCCA model from Thielen and colleagues~\cite{thielen2021} to incorporate envelope information. This is a crucial step, because a 1 in the code does not necessitate that there was audio in the stimulus. By incorporating the envelope and combining these with the events in the structure matrix, it can be avoided that an event is expected even though there was an audio amplitude of zero in the speech signal at that time. This is achieved by element-wise multiplying the event matrix $\mathbf{E}_i$ by the amplitudes of the envelope $\mathbf{A}_i$, before mapping the event matrix to a Toeplitz-like structure matrix. 

Let's assume we have a training dataset $\{(\mathbf{X}_1, y_1), (\mathbf{X}_j, y_j) \dots, (\mathbf{X}_J, y_J)\}$ including the labeled EEG data for $j \in \{1, ..., J\}$ trials with the EEG data $\mathbf{X} \in \mathbb{R}^{C \times T}$ of $C$-many channels and $T$-many samples and the associated binary label $y \in \{0, 1\}$. To find the optimal spatial filter $\mathbf{w}$ and temporal response vector $\mathbf{r}$, a CCA maximizes the correlation $\rho$ in the projected spaces:
\begin{equation}\label{eq:rcca_fit}
    \underset{\mathbf{w}, \mathbf{r}}{\arg\max}~\rho (\mathbf{w}^{\top}\mathbf{S}, \mathbf{r}^{\top}\mathbf{D}) 
\end{equation}
where $\mathbf{S} = [\mathbf{X}_1, \mathbf{X}_j, \dots, \mathbf{X}_J]$ are the concatenated single trials and $\mathbf{D} = [\mathbf{M}_{y_1}, \mathbf{M}_{y_j}, \dots, \mathbf{M}_{y_J}]$ are the concatenated accompanying structure matrices. 

To classify new data $\mathbf{X} \in \mathbb{R}^{C \times T}$ (here $T=\tau$ is the decision window length), rCCA maximizes the correlation $\rho$ between the spatially filtered data and the projected structure matrix that contains the speech envelope:
\begin{equation}\label{eq:rcca_predict}
     \hat{y} = \underset{i}{\arg\max}~\rho (\mathbf{w}^\top\mathbf{X}, \mathbf{r}^\top\mathbf{M}_i)
\end{equation}
In this work, for rCCA, we only used the first CCA component for classification similar to the application of rCCA in the visual domain~\cite{thielen2021}.

The code for the reconvolution CCA approach is available at \url{https://github.com/thijor/pyntbci}.


\mysection{results}

\begin{figure*}[h]
    \centering
    \includegraphics[width=\linewidth]{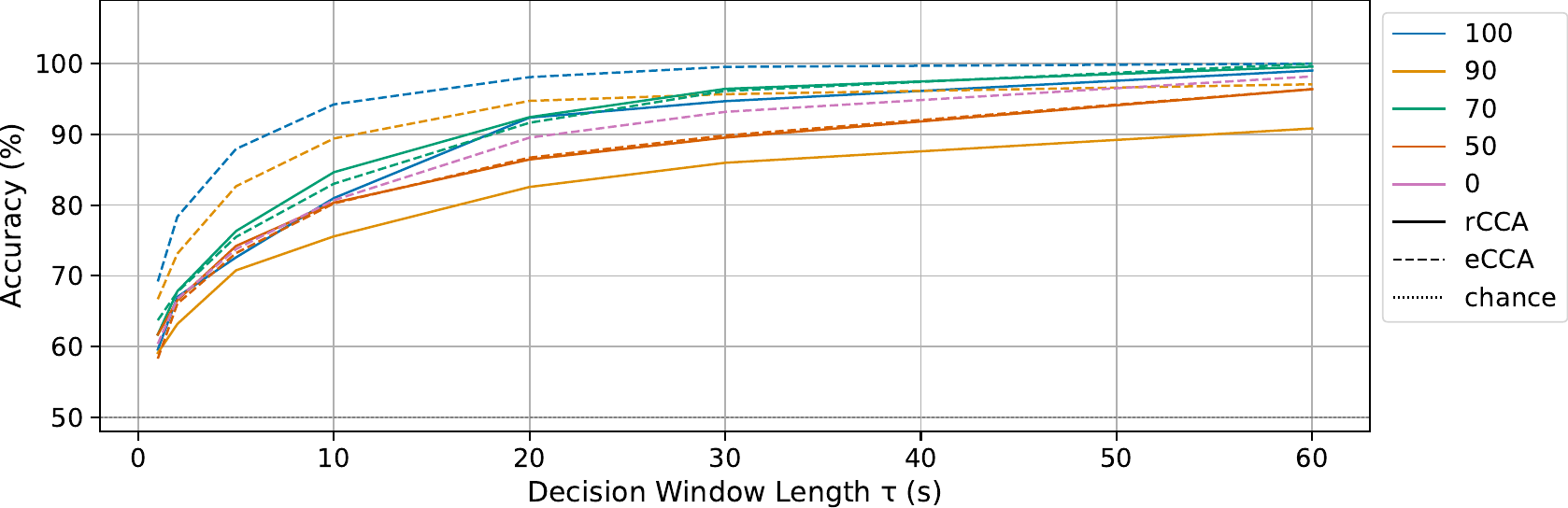}\caption{\label{fig:rcca_ecca_accuracies_contrast}\textbf{Decoding accuracy across decision window length and modulation depth.} Depicted is the grand average classification accuracy across decision window length $\tau$. Colored lines represent the five modulation conditions: 100 (blue), 90 (orange), 70 (green), 50 (red), and 0 (pink). Solid lines show the performance of rCCA and dashed lines for eCCA. The dashed horizontal gray line depicts theoretical chance level accuracy (50\%).}
\end{figure*}

This work aimed to investigate fundamental insights in the application of noise-codes for auditory attention decoding. Two different methods were studied: eCCA which used the speech envelope; and rCCA which leveraged the noise-codes. In total, five conditions were used, from audio without modulation (0), to those with increasing modulation depths (50, 70, 90), to audio with full amplitude modulation (100). To investigate the speed of the models, sliding decision windows of length $\tau$ ranging from 1 to 60\,s were used during testing. The mean classification accuracy for all conditions and both models per decision window length is shown in Fig.~\ref{fig:rcca_ecca_accuracies_contrast}. For an overview of the decoding accuracy for several decision window lengths, see Tab.~\ref{tab:mean-accuracies}.

\begin{table}[h]
\begin{small}
\caption{\label{tab:mean-accuracies}\textbf{Classification accuracy.} Listed are the grand average accuracy for the modulation conditions and both methods at decision window lengths $\tau$ of 1, 10, 30, and 60\,s. Bold values indicate which method (eCCA or rCCA) reached a higher absolute accuracy within that decision window for a particular condition. }
\begin{tabular*}{\columnwidth}{cc@{\extracolsep{\fill}}cccc}
\toprule
\multicolumn{1}{r}{} & \multicolumn{1}{r}{$\tau$} & 1\,s & \multicolumn{1}{c}{10\,s} & \multicolumn{1}{c}{30\,s} & \multicolumn{1}{c}{60\,s} \\  \hline
\midrule
0 & \multicolumn{1}{c|}{eCCA} & 60.4 & 80.7 & 93.2 & 98.2 \\ \hline
\multirow{2}{*}{50} & \multicolumn{1}{c|}{eCCA} & 58.3 & 80.2 & \textbf{89.9} & 96.4 \\
 & \multicolumn{1}{c|}{rCCA} & \textbf{61.7} & \textbf{80.4} & 89.6 & 96.4 \\ \hline 
\multirow{2}{*}{70} & \multicolumn{1}{c|}{eCCA} & \textbf{63.7} & 83.0 & 96.1 & \textbf{100.0} \\
 & \multicolumn{1}{c|}{rCCA} & 61.7 & \textbf{84.7} & \textbf{96.4} & 99.6 \\ \hline
\multirow{2}{*}{90} & \multicolumn{1}{c|}{eCCA} & \textbf{66.7} & \textbf{89.4} & \textbf{95.7} & \textbf{97.1} \\
 & \multicolumn{1}{c|}{rCCA} & 59.1 & 75.6 & 86.0 & 90.8 \\ \hline 
\multirow{2}{*}{100} & \multicolumn{1}{c|}{eCCA} & \textbf{69.2} & \textbf{94.2} & \textbf{99.6} & \textbf{100.0} \\
 & \multicolumn{1}{c|}{rCCA} & 59.6 & 81.0 & 94.7 & 99.0 \\ \bottomrule
\end{tabular*}
\end{small}
\end{table}

The eCCA method applied to the unmodulated condition (0), which was the baseline in this study, reached a 60\,\% decoding accuracy for a 1\,s decoding window, 80.7\,\% for 10\,s, 93.2\,\% for 30\,s, and 98.2\,\% for 60\,s. Also using eCCA, but with the 100 modulation, the highest absolute decoding performance was reached for all decision windows; 69.2\,\%, 94.2\,\%, 99.6\,\%, and 100.0\,\% for 1, 10, 30, and 60\,s respectively. Modulation condition 90 performed better than 0 for all decision window lengths except 60\,s, where it reached 97.1\,\%. The 70 modulation performed better than 0 modulation for all decision windows. Lastly, modulation condition 50 reached lower performances than 0 modulation for all decision windows. 

When using rCCA and comparing absolute values, it was observed that the 70 modulation condition performed best for all decision window lengths, whereas 90 modulation performed the worst. The 100 modulation performed worse than 70 modulation but overall better than 50 modulation. For example at $\tau=30$\,s and ordered at increasing accuracy, 90 modulation reached 86.0\,\%, 50 modulation 89.6\,\%, 100 modulation reached 94.7\,\%, and 70 modulation achieved a performance of 99.6\,\%.

When comparing the absolute performance values for eCCA against rCCA, it can be observed that the 100 and 90 modulation conditions reached higher performances with eCCA, for all decision windows. For modulation conditions 70 and 50, rCCA reached overall on par performances as eCCA.

When comparing the absolute performance values of eCCA on the unmodulated condition (0) and the modulation conditions using rCCA, it can be observed that the 70 modulation condition (with rCCA) achieved higher performances than the unmodulated condition (with eCCA) for all decision window lengths. For example, for \mbox{$\tau=10$\,s}, 80.7\,\% was reached for 0 modulation with eCCA and 84.7\,\% for 70 modulation with rCCA, and for $\tau=30$\,s 93.2\,\% and 96.4\,\%, respectively. The 100 modulation with rCCA reached higher performances than 0 modulation for decision window lengths of $\tau\geq10$\,s. The 90 modulation condition with rCCA performed worse than 0 modulation with eCCA for all decision window lengths. For the 50 modulation condition with rCCA, the 0 modulation with eCCA reached higher performances, although for some decision windows rCCA reached an on par or somewhat higher performance, for example for $\tau=1$\,s 50 modulation with rCCA reached 61.7\,\% and 0 modulation with eCCA 60.4\,\%. 

In general, using longer decision window lengths $\tau$ was beneficial for the decoding performance of both methods for all conditions.


\mysection{discussion}

This preliminary work aims to contribute to our understanding of fundamental protocol design decisions when noise-tagging is integrated to perform AAD. We argue that noise-tagging may provide additional information to the audio which could enhance AAD performance. In this pilot study with sequential presentation of the stimuli, we studied five modulation conditions for the speech signals and compared two CCA-based decoding approaches. The results showed that the envelope CCA (eCCA) method achieved higher performances with full modulation than without modulation, while the alternative reconvolution CCA (rCCA) preferred a 70 percent modulation intensity to reach peak performance.

For all decision window lengths, the 100 and 70 modulation conditions performed better than the unmodulated condition for eCCA. It could be speculated that this is due to the modulation adding distinctive uncorrelated high-frequency content. Generally, a 1-9\,Hz range was found most informative for cortical tracking of speech envelopes~\cite{das2016}. Arguably, adding the noise-tags at a 40\,Hz bitrate increases the envelope's frequency range, which could increase the speed with which the envelopes become distinctive for eCCA. Therefore, we used a 20\,Hz lowpass and a higher sampling frequency.

The lowest modulation depth that reached on par or higher performance with the unmodulated condition was a modulation of 70, both for using eCCA and rCCA. For both methods, a 50 modulation depth was able to reach similar performances as the 0 modulation, for decision window lengths of $\tau\leq10$\,s. Future work could assess the perception thresholds for modulation depths, to obtain the least intrusive protocols for high usability.

Overall, the performance of rCCA matched that of eCCA, or was lower. However, differences exist between these methods. Firstly, eCCA emphasizes global and higher-order activity associated with the speech envelope, while rCCA may focus more on early sensory responses evoked by the noise-codes. Secondly, eCCA uses multiple CCA components and an LDA classifier. While such enhancements could potentially also benefit rCCA, we stayed close to existing literature for this pilot study. Future investigations should evaluate rCCA's decoding performance upon integrating these additional optimizations.

A strong characteristic of applying amplitude modulation using noise-codes is that the resulting decoding is less limited by the distinctiveness of the envelope of the audio signal, for example in speech. The method could therefore also be applied more broadly to any type of audio signal, the envelope of which may be more or less correlated, such as music. 

As this work is a preliminary study, a number of inherent limitations should be noted. First, the experiment only included five participants, of which two were authors with substantial experience as BCI users. This is a small sample size, which included motivated participants, and needs to be enlarged in future studies. Second, in order to assess the feasibility of decoding a c-AEP response, this pilot used sequential stimulation instead of presenting the two speakers in parallel. Future studies need to investigate, if our observations generalize also to parallel stimulus presentation protocols.

Future work could include various other improvements, such as the noise-codes optimized to maintain speech qualities. Additionally, the noise-codes could be shortened, as the number of classes typically may be lower than 63 in an auditory attention scenario.

\mysection{conclusion}

Our work showed that adding noise-tags to a speech signal in a sequential paradigm can enhance the decoding performance compared to decoding the original unmodulated speech signal. Specifically, for shorter decision window lengths all higher modulation depths (100, 90 and 70) performed better than the unmodulated condition. The rCCA method on the 70 modulation condition also performed better than the unmodulated condition. Lastly, for the modulated conditions, the eCCA method performed better than or on par with the rCCA method. Overall, these results show the potential of using noise-tags in the auditory modality, and is the first step towards using the noise-tagging protocol for auditory attention decoding. 

\mysection{acknowledgments}

We thank K. van der Heijden for helpful insights into AAD.


\mysection{references}
\printbibliography[heading=none]

\end{document}